\documentclass[prd,twocolumn,showpacs,preprintnumbers]{revtex4}
\pdfoutput=1
\usepackage{amsfonts,amsmath,amssymb}
\usepackage{graphicx}
\usepackage{color}
\usepackage{natbib}
\pdfoutput=1

\begin{document}

\newcommand{\apjl}{Astrophys. J. Lett.}
\newcommand{\apjs}{Astrophys. J. Suppl. Ser.}
\newcommand{\aap}{Astron. \& Astrophys.}
\newcommand{\aj}{Astron. J.}
\newcommand{\araa}{Ann. Rev. Astron. Astrophys. } 
\newcommand{\mnras}{Mon. Not. R. Astron. Soc.}
\newcommand{\jcap}{JCAP}
\newcommand{\pasj}{PASJ}
\newcommand{\pasa}{Pub. Astro. Soc. Aust.}

\title{The Galactic One-Way Shapiro Delay to PSR B1937+21}
\author{S. \surname{Desai}$^{1}$} \altaffiliation{E-mail: shntn05@gmail.com}
\author{E.O. \surname{Kahya}$^2$} \altaffiliation{E-mail: eokahya@itu.edu.tr}

\affiliation{$^{1}$Excellence Cluster Universe, Boltzmannstrasse 2, D-85748 Garching, Germany}
\affiliation{$^{2}$Department of Physics, Istanbul Technical University, Maslak 34469 Istanbul, Turkey}

\begin{abstract}

The  time delay experienced by a light ray as it passes through a changing gravitational potential  by  a non-zero mass distribution along the line of sight is usually referred to as Shapiro delay.  Shapiro delay  has been extensively measured in  the Solar system and in binary pulsars, enabling stringent   tests of general relativity as well as measurement of  neutron star masses .  However, Shapiro delay is ubiquitous and  experienced by all astrophysical messengers on their way from the source to the Earth. We calculate the ``one-way'' static Shapiro delay for the first discovered millisecond pulsar PSR~B1937+21, by including  the  contributions from both  the dark matter  and baryonic matter between this pulsar and the  Earth. We find a value of approximately 5 days (of which 4.74 days is from  the dark matter  and 0.22 days from the baryonic matter).  We also calculate the modulation of Shapiro delay from  the motion of a  single dark matter halo,  and also evaluate the cumulative effects of  the motion of matter distribution on the change in  pulsar's period and its derivative. The time-dependent effects are too small  to be detected with the current timing noise observed for this pulsar. Finally, we would like to emphasize that although the one-way Shapiro delay is mostly of academic interest for  electromagnetic astronomy, its ubiquity should not be forgotten in the era of multi-messenger astronomy.
\end{abstract}

\maketitle

\section{Introduction}

In 1964, I. Shapiro~\cite{Shapiro} calculated and then measured~\cite{Shapiro66} the  round-trip time delay of a radar signal to the inner planets of our solar system, which is caused by the  gravitational field of the Sun. This delay is known as    ``Shapiro delay'' and has been measured precisely in the solar system over the last five decades, allowing very stringent tests of  general relativity (GR) and in particular the PPN $\gamma$ parameter~\cite{Will}. The current best solar system constraints come  from the Cassini mission, which agree with the GR prediction to within 
$10^{-5}$~\cite{Cassini}.  The calculation   of Shapiro delay  has also  been generalized for  a time-varying gravitational field~\cite{Hellings,Kopeikin99,Kopeikin01} and  experimentally confirmed to agree with the predictions  of GR~\cite{Kopeikin01,Kopeikin03}. It has also been calculated for some alternate theories of gravity in anticipation of future solar system 
based satellite experiments~\cite{Asada08,Carlip04}.
Besides its use as a test of GR, Shapiro delay has been routinely used as an astrophysical  probe to measure the masses of pulsars in binary systems~\cite{Demorest,Corongiu12,Lohmer05}, allowing us to  constrain   the neutron star equation of state.  Shapiro delay has been proposed  as the possible cause  of the low frequency noise in timing residuals of pulsars in globular clusters~\cite{Kopeikin05}. This delay is also one of the contributing factors for   the observed time delays between multiply   lensed images from quasars ~\cite{Blandford}.  These time delays have been measured and used to constrain Hubble  constant and other  cosmological parameters~\cite{Kochanek}.

In this paper, we would like to focus on another facet of Shapiro delay, which is the total delay experienced by any astrophysical messenger
from cosmic rays to gravitational waves on its way to the Earth from the source, due to the gravitational potential of all the intervening 
mass distribution along the line of sight. We refer to this as the ``one-way'' Shapiro delay. Although this was first calculated in 1988~\cite{Longo,Krauss} 
for the gravitational potential of our galaxy, it is  rarely mentioned in astrophysical literature. This  is because one can never measure the absolute value
of this delay and only its time-dependence has observational consequences, in case the source is a steady-state emitter. However,  one can measure the difference in
the static component of the Shapiro delay  between photons and neutrinos/gravitational waves in the case of simultaneous detection of multiple cosmic messengers, and when we know the relative departure time at the source. In this situation, although the absolute static component of the delay can never be measured, we can use one astrophysical messenger as a clock to time the other one and thereby test the  equivalence principle for the non-electromagnetic messenger.

However, we would like to enumerate some  examples  of various astrophysical measurements, which have  already been done or planned in the future, for which  one-way Shapiro delay plays a central role in the final results which are derived from these observations.
As a strawman, we then focus on the Shapiro delay calculation for one astrophysical source, which is the millisecond pulsar PSR B1937+21. We calculate the total Shapiro delay from both the dark matter and baryonic matter assuming static spherically symmetric geometry. We then estimate the time-dependent corrections on the pulsar period and its derivative due to the velocity of the matter distribution along the line of sight and then discuss the prospects
for detection.

The outline of this paper is as follows. In Sect.~\ref{sec:psr}, we provide a brief history of one-way Shapiro delay calculations in literature, including measurements from SN1987A and implications for fundamental physics from these,  as well as  point out other astrophysical examples where one-way Shapiro delay is relevant.
We then set up our formalism for calculating the static Shapiro delay from PSR B1937+21 in Sect.~\ref{shapirodelay}, including the mass models used for dark matter (Sect.~\ref{sec:dm}) and baryonic matter (Sect.~\ref{sec:bm}), and then calculate the total delay assuming a static potential. In Sect.~\ref{sec:shapirov},
 we calculate the corrections due to the velocity of the matter distribution.  We then conclude in Sect.~\ref{sec:conclusions}.

\section{One-Way Shapiro Delay}
\label{sec:psr}

The first paper  which explicitly mentions  the existence  of one-way Shapiro delay from our galaxy (to the best of our knowledge) is by Backer and Hellings ~\cite{Hellings86} (see their Eqn.~4.3), which was  written  in the context of pulsar timing observations.  However, they argued that since  the galactic gravitational potential is essentially static  over a ten year period, there is no need to model for it. Following the detection of neutrinos from SN1987A~\cite{IMB,Kamioka}, it was pointed out  that  the neutrinos from SN1987A also experience this  one-way Shapiro delay due to the gravitational potential of the intervening matter along the line of sight. Two independent groups in back to back papers calculated the delay by modeling the gravitational field of the   Milky Way
as a point mass. The value for the one-way delay ranged from  one to six  months for different models of the  galactic gravitational potential~\cite{Longo,Krauss}.  This delay calculation was also generalized for a non-zero neutrino mass~\cite{Bose} and the difference due to the neutrino mass was shown to be negligible for the neutrino energies detected from SN1987A.  The near-simultaneous arrival of photons and neutrinos from this core-collapse supernova confirmed that Shapiro delay for neutrinos is same as that for photons to within 0.2-0.5\%~\cite{Longo,Krauss}. To date, this is the only  direct observational evidence we have that neutrinos are affected by gravity and obey the weak equivalence principle. These observations also constrained the difference in relative couplings of gravitational interactions of matter and anti-matter to within $10^{-5} - 10^{-6}$~\cite{Losecco}.  We should point out that  the calculation of one-way Shapiro delay for sources in our galaxy and local neighborhood is decoupled from measurement of distances, and if the gravitational potential along the line of sight changes, then so would the arrival time.

The one-way Shapiro delay is also a  very important factor in searches for gravitational waves from sources with electromagnetic counterparts, which 
should be expected with  advanced LIGO commencing operations in Sept. 2015. In case gravitational waves arrive at the same time as photons, it would be the first direct evidence 
that gravitational waves gravitate, or that ``gravity begets gravity''. We  could also use measurements of relative Shapiro delay between gravitational waves and photons/neutrinos  to confirm  or rule out modified theories of gravity designed to explain the dark matter conundrum~\cite{Kahya08,Desai,Kahya10}.

However, besides the above examples the galactic one-way Shapiro delay due to the gravitational potential of our galaxy is hardly ever discussed in literature. 
This could be because for all practical purposes, this delay is only of academic interest, since it is much smaller than the vacuum light travel time. Other
possible reasons are that  one can never know when a photon left an astrophysical source and the gravitational potential of our galaxy changes very slowly. The measurement of one-way Shapiro delay has practical relevance only in case of two independent messengers seen from an astrophysical source, and for which we know the delay between them at the source. To date, the only such example Mother Nature has provided us, is the simultaneous detection of neutrinos and photons from SN1987A. However with the recent detection of high energy astrophysical neutrinos from IceCube~\cite{icecube}, and  the expected detection of gravitational waves now that advanced LIGO has started taking data, we hope that the ubiquity of one-way Shapiro delay is  not forgotten.

We now point out the relevance of one-way Shapiro delay for some more astrophysical measurements. In anticipation of expected gravitational
wave observations from sources with electromagnetic and neutrino counterparts, there have been many proposed tests of various fundamental
physics parameters such as vacuum speed of gravitational waves, mass of the graviton,  mass of the neutrino.  
~\cite{Larson13,Nishizawa15,Arnaud01}. A key assumption for all such measurements (even though its rarely stated)  is that the one-way Shapiro delay is the same for gravitational waves and neutrinos/photons. However, we don't know  as of now whether gravitational waves follow the same geodesics (as photons) from the intervening matter distribution, because they have not been directly detected yet.  In some alternate gravity theories which violate the strong equivalence principle, they do not~\cite{Desai}.
 In purely electromagnetic astronomy, the equality of one-way Shapiro delay along different lines of sight is also a key assumption in the observations of light echoes from distant supernovae~\cite{Rest}.

Therefore, even though the general lore is that one-way Shapiro delay is only of academic interest,  the above examples show  some applications of  Shapiro delay observations in both astrophysical and fundamental physics measurements, and especially for multi-messenger astronomy.

\section{Shapiro Delay for PSR B1937+21}
\label{shapirodelay}
We now turn to the delay calculation for the pulsar PSR B1937+21. This is the first ever  discovered millisecond pulsar,  which  has a rotational period of  1.55 milliseconds~\cite{Kulkarni82}, dispersion measure of about 71~pc $\mathrm{cm^{-3}}$, and  timing
residuals of about 2 $\mu s$ from over three decades of observations~\cite{Kaspi}.

The total light travel time from this pulsar  includes the geometric propagation delay due to   distance and proper motion of the pulsar,   the one-way Shapiro delay from  all intervening masses along the line of sight~\cite{Hellings86,Manchester}), as well as two additional frequency-dependent delay terms  from propagation in the interstellar medium due to dispersion and birefringence~\cite{Cordes13}. However,
these frequency dependent terms are small compared to the light travel time.
We also note that the   Shapiro delay calculation from the inner solar system planets is included  in the {\tt TEMPO2} software, which is routinely used in  the analysis of pulsar data~\cite{Manchester}. We now calculate the total one-way Shapiro delay for this pulsar from the static gravitational potential of the dark matter, and also the  modulations to the static Shapiro delay from the motion  of dark and baryonic matter.

\subsection{Formalism for Calculating the one-way Shapiro Delay}
\label{sec:shapiro}
We discuss the details of the Shapiro delay calculation  experienced by the radio wave on its way from the pulsar
to the Earth. Previously, we have calculated this using only the dark matter contributions~\cite{Desai} for a few selected astrophysical sources (SN1987A, GRB070201, and Sco-X1), and  also as a function of distance in our galaxy~\cite{Kahya10}. In this paper, we do this calculation for PSR B1937+21 from both the dark  matter and baryons  along 
the line of sight. We briefly review the formalism for calculating the static Shapiro delay, which follows the same method  and  notations as in ~\cite{Desai,Kahya10} and the reader can refer to these papers for more details.

To calculate the delay, we shall assume  a static spherically symmetric geometry, which is a good approximation for dark matter models of the Milky Way.   The Schwarzschild line element for this geometry is given by

\begin{equation}
ds^2 \equiv -B(r) c^2 dt^2 + A(r) dr^2 + r^2 d\Omega^2 \; . \label{ds2}
\end{equation}

We can calculate $A(r)$ and $B(r)$ as a function of  the mass distribution  by solving Einstein's equations. For pressure-less dust, they take the form:

\begin{eqnarray}
\frac{B}{A} \Biggl[ \frac{A'}{r A} + \Bigl(\frac{A-1}{r^2}\Bigr)\Biggr]
& = & \frac{8\pi G}{c^2} \, \rho \; , \label{eq:3} \\
\frac{B'}{r B} - \Bigl(\frac{A-1}{r^2}\Bigr) & = & 0 \; . \label{eq:4}
\end{eqnarray}

We assume that the deviations from flat geometry are small and therefore 
$A(r)$ and $B(r)$ can be written as 
\begin{equation}
A(r) \equiv 1 + \Delta A(r) \qquad , \qquad B(r) \equiv 1 + \Delta B(r) \; .
\end{equation}

Linearizing eqns.~(\ref{eq:3}-\ref{eq:4}) and solving these two equations  give the following
expressions for  $\Delta A(r)$ and $\Delta B(r)$ in terms of density profiles or mass functions.
\begin{equation}
M(r) \equiv 4 \pi \int_0^r r^{\prime 2} dr' \, \rho(r') \; .
\end{equation}

The linearized solution of Eqn.~\ref{eq:3} is 

\begin{equation}
\Delta A(r) = \frac{8 \pi G}{c^2 r} \int_0^r dr' \, r^{\prime 2} \rho(r') =
\frac{2 G}{c^2} \, \frac{M(r)}{r} \; , \label{DA}
\end{equation}

\begin{equation}
\Delta B(r) = -\!\!\int_r^{\infty} \!\!\!\!\! dr'  \, \frac{\Delta A(r')}{r'} .
\label{DB}
\end{equation}

After rewriting the metric in terms of the mass distributions, one can calculate
Shapiro delay by solving the null geodesic equation . The total Shapiro delay ($T_{\rm shapiro}$)
of a photon with its initial Cartesian $(\vec{x}_1)$ and final positions $(\vec{x}_2)$  is  now given by~\cite{Desai} :
\begin{eqnarray}
\lefteqn{c  T_{\rm shapiro} = \frac{\Delta \vec{x} \cdot \vec{x}_1}{2 \Delta x} \,
\Delta B(r_1) - \frac{\Delta \vec{x} \cdot \vec{x}_2}{2 \Delta x} \,
\Delta B(r_2) } \nonumber \\
& & \hspace{0.5cm} + \int_{r_2}^{r_1} \!\!\! dr \, \frac{2 G M(r)}{c^2 r}
\sqrt{1 - \frac{r_1^2 \Delta x^2 - (\vec{x}_1 \cdot
\Delta \vec{x})^2}{r^2} } \; , \qquad \label{cdt}
\label{eq:shapiro}
\end{eqnarray}

where $r_1$ and $r_2$ denote the initial and final radial position of the photon, $\Delta x = \lvert \vec{x}_2 - \vec{x}_1 \rvert$.
We now evaluate these integrals numerically for a given dark matter and baryon
distribution. But before  doing that, we make  an  order of magnitude estimate  of the two
contributions.  $\Delta A(r)$ and $\Delta B(r) $ can be approximated as follows:

\begin{equation}
\Delta A(r)  \approx \; \epsilon \;+  \; \epsilon_{*} \; , \; \Delta B(r)  \approx  - \;\epsilon \;+  \; \epsilon_{*} \ln(\frac{r}{r_s\
}) .
\label{eq:5}
\end{equation}  

where $\epsilon$ is the usual Schwarzschild term $\equiv (2GM/c^2r$), where $M$ is the baryonic mass,  and  $\epsilon_{*}$ is  due  to the dark matter  contributions given by $\epsilon_{*} \equiv 2v_*^2/c^2$, where $v_*$ is the asymptotic rotation speed of the  Milky Way.
If we assume  that the baryonic mass at radial distance of 10 kpc to be of order $5 \times 10^{10} M_{\odot}$, one can get a back of the envelope estimate  of the magnitude for $\epsilon$ to be $10^{-6}$ and assuming $v_*$  to be 200 km/sec, then $\epsilon_{*}$  is of the order $6 \times 10^{-7}$. Therefore both dark matter and baryonic effects are of the same order of magnitude and should be taken into account while calculating the Shapiro delay. We now proceed to more
detailed calculations.

\subsection{Shapiro Delay from Dark Matter}
\label{sec:dm}
Over the last decade there has been a lot of effort to model the dark matter halo of the Milky Way,
using different observational tracers~\cite{Rix}.   
For this paper, we shall  use the  Milky Way dark matter  profile from Klypin et al~\cite{Klypin}. Currently, the 
uncertainty in dark matter mass for our Milky Way Halo is about 30\%~\cite{Rix}. However, this uncertainty should be reduced  using 
results from the GAIA satellite.

We briefly review Klypin et al's posited dark matter profile and its associated parameters~\cite{Klypin}. 
They assumed a Navarro-Frenk-White (NFW)~\cite{nfw} profile given by  : 
\begin{eqnarray}
\label{eq:NFWa}
        \rho_{\rm halo}(r) &=& \frac{\rho_s}{x(1+x)^2}, \quad x = r/r_s \\
        M_{\rm halo}(r) &=& 4\pi\rho_sr_s^3f(x) \\ &=& M_{\rm vir}f(x)/f(C), \\
        f(x) &=& \ln(1+x) -\frac{x}{1+x},\\  C &=& r_{\rm vir}/r_s,\\
\label{eq:NFWz}
\end{eqnarray}
\noindent where $C$ and $M_{\rm vir}$ are the halo concentration and
virial mass, and $r_{\rm vir}$ is the virial radius. Details about the other
terms can be found in~\cite{Klypin}.
Following this paper, we assume  $r_{\rm vir}=258$ kpc, $M_{\rm vir}=10^{12} M_{\odot}$ and $C=12$.
Using these values, we can calculate $\rho_s \simeq 0.186~{\rm GeV/cm^3}$.
Therefore, using this value of  $\rho_s$ and assuming an NFW profile (given in Eqn.~\ref{eq:NFWa}), we
can calculate $\Delta A(r)$ and $\Delta B(r)$ at any distance ($r$)  using Eqns.
~\ref{DA} and \ref{DB} respectively. The dark matter induced Shapiro delay can 
then be evaluated using  Eqn.~\ref{cdt}. 

\begin{figure}[htb]
\includegraphics[width=.5\textwidth]{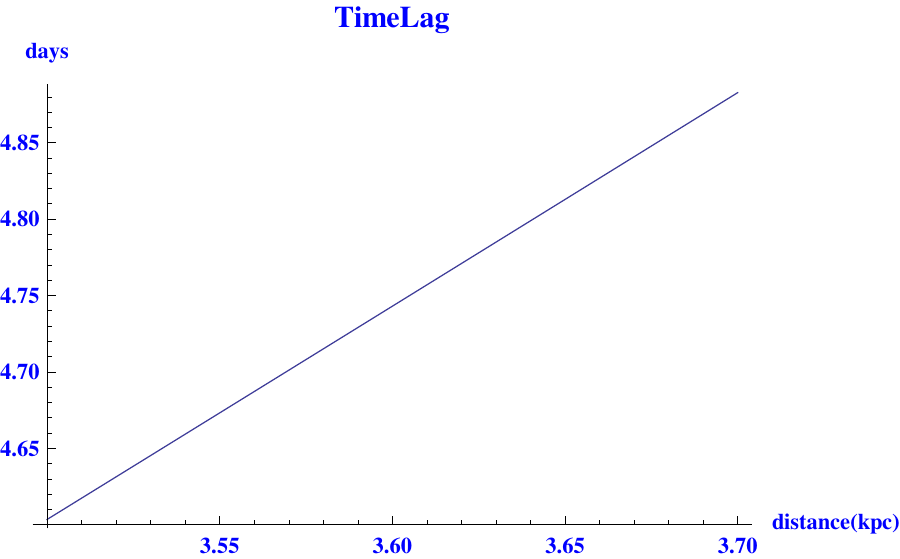}
\caption{Dark matter induced Shapiro delay to PSR B1937+21 as a function of distance in the vicinity of the pulsar.}
\label{shapirodistance}
\end{figure}

\begin{figure}[htb]
\includegraphics[width=.5\textwidth]{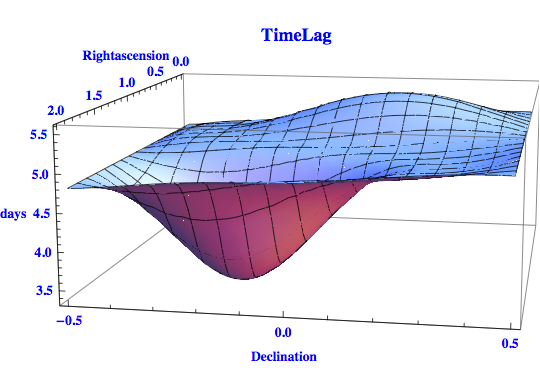}
\caption{Dark matter induced Shapiro delay to PSR B1937+21   as a function of Right Ascension and declination near the pulsar. The Right Ascension and declination are shown in radians.}
\label{shapirospatial}
\end{figure}

We now do these numerical integrations for PSR B1937+21. For the distance to the pulsar
and its position on the sky, we use the tabulated values from  Table I in Nicastro et al~\cite{Nicastro}.
From this paper, we also note that the lower limit on the distance of the pulsar is  $\simeq$ 3.6 kpc 
with an upper limit of 21 kpc. This pulsar is located at Right Ascension of 19hr~39mt~38.5~sec and declination of $21^{\circ} 34^{'} 59^{''}$.
For the calculations in this  paper, we shall assume that the pulsar is at a distance of 3.6 kpc.
The total dark matter mass between the Earth and the pulsar for the above parameters is 
$1.8 \times 10^{10} M_{\odot}$.
Given this spatial location and distance, the calculated dark matter induced Shapiro delay is shown 
below as a function of distance and celestial coordinates near PSR B1937+21 in Figs.~\ref{shapirodistance} and ~\ref{shapirospatial} 
respectively. Therefore for this dark matter profile, we find that the dark matter induced Shapiro delay is approximately
 4.74 days.

\subsection{Shapiro Delay from Baryonic Matter}
\label{sec:bm}
To  calculate the Shapiro delay from baryonic matter, we assume that the total baryonic matter is 
given by the sum of bulge and disk components, and assume spherical symmetry for mass distributions of both the bulge and the disk. We use the mass models from Xue et al~\cite{Xue}, which
assumed a Hernquist profile~\cite{Hernquist} for the galactic bulge with total mass given by
 $M_{\rm bulge} = 1.5 \times 10^{10} M_{\odot}$,  and the Miyamoto \& Nagai profile~\cite{Miyamoto75,Smith07} for the disk,  with total mass given by 
 $M_{\rm disk} = 5 \times 10^{10} M_{\odot}$. Therefore the total mass is equal to $6.5 \times 10^{10} M_{\odot}$.
We should however point out that estimates for the total disk and bulge mass of the Milky Way differ  a lot in literature. For example,  using measurements of the stellar luminosity function mass of the bulge and disk were estimated to be,  $M_{\rm bulge} = 1.3 \times 10^{10} M_{\odot}$ and $M_{disk} = (4.9-6.7) \times 10^{10} M_{\odot}$~\cite{Flynn}. Therefore, in this model the total baryonic mass they obtain is  $(7.1 \pm 0.9) \times 10^{10} M_{\odot}$, of which $(4.9 \pm 0.4) \times 10^{10} M_{\odot}$ lies within the solar circle. Some other estimates of bulge mass are about twice this value.  From the DENIS near-infrared large scale survey, the mass of bulge is assumed to be $(2.4 \pm 0.6) \times 10^{10} M_{\odot}$~\cite{Picaud}.
However, since we are interested in an order of magnitude estimate we use the bulge and disk mass models from Xue et al~\cite{Xue}. Using this value for the total baryonic mass, we can calculate $\Delta A(r)$ and $\Delta B(r)$ directly and hence the Shapiro delay in the same way as for the dark matter.  After doing this,   we get a value of 0.13 days for the bulge and 0.086 days for the disk. The total Shapiro delay from baryonic matter along the line of sight is approximately 0.22 days and is negligible compared to the dark matter contributions.

Therefore, the total galactic Shapiro delay to PSR B1937+21 by summing both the baryonic and dark matter contributions is  about five days. Since the purpose of this calculation is only for pedagogy, we do not calculate any systematic  errors for this delay.

\subsection{Velocity Dependent Corrections to Shapiro Delay}
\label{sec:shapirov}
Since only the  modulations to the static Shapiro delay are potentially detectable, we do a feasibility study of the detection prospects
by estimating its time dependence.
We first calculate the  corrections to the static Shapiro delay estimated in Sect.~\ref{sec:shapiro} by the motion of matter along the line of sight between the Earth and the pulsar. For an elaborate calculation, we would need to linearize the Kerr metric and use the number density of  dark matter haloes from theories of structure formation, along with their phase space velocity distribution to calculate the velocity dependent corrections to the static Shapiro delay.  Here, we use simplified assumptions to calculate order of magnitude effects of the Shapiro delay induced modulations. We first do the calculation for a single
gravitating dark matter mass close to the line of sight between the Earth and the pulsar (as this is the signal most likely to be detected) and then calculate the cumulative stochastic effect from all the masses at a given instant.

The effects of modulation on the Shapiro delay from dark matter for both pulsars in our galaxy and globular clusters and their observational signatures have been investigated by a number of authors~\cite{Larchenkova95,Fargion,Walker,Baghram,Fry,Ishiyama,Larchenkova07,Pshirkov08}. Besides  Shapiro delay, orbiting dark matter haloes  could also induce gravitational perturbations~\cite{Seto,Pshirkov,Ishiyama},  and frequency-dependent modulations due to Doppler effects~\cite{Baghram}, all of which could affect the timing residuals in millisecond pulsars.

To  calculate the dependence on the velocity of the mass along the line of sight,  we use the expressions from  Refs.~\cite{Larchenkova07,Walker}.
The starting point in these calculations is a generalization of the standard textbook formula for Shapiro delay for a point mass given
by~\cite{Shapiro} :
\begin{equation}
T_{shapiro} = -{2GM \over c^3} \ln(1-R\cdot x) , 
\end{equation}
where $T_{shapiro}$ is the total Shapiro delay for a point mass,  $M$ is the mass of the object which causes the delay, $R$ is the unit vector from the Earth to the pulsar and $x$
is the unit vector from the Earth to the gravitating mass. To calculate velocity dependent modulations, we calculate the change in
$R\cdot x$, caused by the motion of the gravitating mass, and then compare the delay  assuming the mass has no velocity.
 If we assume that the gravitating mass moves with velocity $v$,  $T_0$ is the time of conjunction,  and the impact parameter $d$ (or the point of closest approach between the pulsar and the mass) is much smaller  than the distance between the Earth and the pulsar, then $R\cdot x$ can be expanded
as a Taylor series in $v/d$ and the lowest order velocity-dependent corrections are given by~\cite{Larchenkova07,Walker}:
\begin{equation}
\Delta T_{vel} = -{2GM \over c^3} \ln(1+\left(v/d\right)^2\cdot(t-T_0)^2) .
\label{eq:velocity}
\end{equation}

where $\Delta T_{vel}$ is the approximate velocity-dependent Shapiro delay correction which must be added to the static part given by Eqn.~\ref{eq:shapiro}. We note that the velocity dependent corrections are opposite in sign to the static component of the Shapiro delay.
Since the proper motion of PSR 1937+21 is small, we assume that $|v|$ is dominated by the Galactic rotation velocity at the position of the Sun or the Local Standard of Rest velocity, and has a value of $\sim$ 220 km/sec in the direction of Galactic Center~\cite{Gondolo}. For the orbiting matter along the line of sight, only those dark matter haloes or
 baryonic clumps   which have small impact parameters will leave an observational imprint on the timing properties of the pulsar signal.  
We first estimate the time variation by a single dark matter halo hovering close to the line of sight between the Earth and pulsar. We choose impact parameters of  $10^{-4}$ pc~\cite{Fry}, since this is the maximum impact parameter, which a dark matter halo could have relative to our line of sight to be of observational significance.
Assuming a median dark matter halo mass of $100~M_{\oplus}$ which is typical of dark matter haloes in our Milky Way~\cite{Loeb}, Eqn.~\ref{eq:velocity} tells us that the peak to peak variation in $\Delta T_{vel}$ with respect to conjunction is about 5~ns over a period of five years. The total number of transits  from all dark matter haloes with such impact parameters can be calculated using the gravitational lensing optical  depth~\cite{Fry} and is  $\sim 0.05$ /year between the Earth and PSR B1937+21. The expected number of transits from baryonic clumps is even smaller. Therefore, we need to monitor the pulsar for  about 25 years  to experience at least one such dark matter halo transit having  the characteristic bell-shaped  signature of Shapiro delay with an 
amplitude of  about 5~ns.  
However, in practice this is much smaller than the timing residuals seen for this pulsar which is $\mathcal{O}(\mu sec)$~\cite{Kaspi} and the dominant sources of timing noise, such as from  pulsar spin-down  are usually fitted  for in the timing analysis.  So the first pre-requisite for detection is that
the timing residuals be reduced by at least three orders of  magnitude.  The  full list of all sources of  timing noise in millisecond pulsars and ways to mitigate or accurately model these terms  to  reduce the residuals to  $\mathcal{O}(nsec)$  during the Square Kilometer array (SKA) era are  discussed in ~\cite{Cordes13,Kramer}.

We now calculate the cumulative effects of all the dark and baryonic matter on the change in  pulsar's period and its derivative at a given 
instant. We assume that the pulsar's true period is $P_0$, and the observed period due to Shapiro-delay induced modulation is $P$, then by taking the derivatives of Eqn.~\ref{eq:velocity}, we obtain the following expressions for the fractional change in the period and its derivative~\cite{Walker}: 

\begin{eqnarray}
\label{eq:Pdot}
\Delta & = &   -{4GM \over c^3} \frac{v}{d} \cos (\theta) , \\ 
\dot{\Delta} & = &   -{4GM \over c^3} \left(\frac{v}{d}\right)^2 \left [1 -2 \cos^2(\theta)\right] .
\label{eq:Pdotdot}
\end{eqnarray}
where $\Delta=\frac{P-P_0}{P_0}$, $\dot{\Delta} \sim \frac{\dot{P} - \dot{P_0}}{P_0}$ (after neglecting $\dot{P_0} \Delta$ term),  and $\cos(\theta)=R \cdot x$. 

If we assume that the dark matter haloes are randomly aligned between the Earth and the pulsar~\cite{Walker}, then $\langle \cos\theta \rangle=0$,  and $\langle \cos^2\theta \rangle=0.5$, and the total effect from summing the terms in Eqn.~\ref{eq:Pdot} will  average out to zero. 
We therefore compute the variance of the above quantities, which  have non-zero values and can be written as~\cite{Walker} :
\begin{eqnarray}
\langle{\Delta}^2\rangle & = &  8 \left(GM \over c^3\right)^2\pi \Sigma v^2 \ln N , \\
\langle\dot{\Delta}^2 \rangle & = &  8 \left(GM \over c^3\right)^2\left(\pi \Sigma v^2\right)^2 .
\end{eqnarray} 
where $\Sigma$ is the surface density of all dark or baryonic matter which contributes to the Shapiro delay, and $N$ is the total number of distinct gravitating objects which contribute to Shapiro delay. In the above equations, the averaging is done over shorter time-scales, over many multiples of the pulsar period. For the dark matter between the Earth and the pulsar, assuming $M=100 M_{\oplus}$, $\Sigma=4 \times 10^5 pc^{-2}$, and $N=6 \times 10^{13}$ we obtain $\langle{\Delta}^2\rangle \sim 1.7 \times 10^{-37}$ and $\langle \dot{\Delta}^2 \rangle \sim 8 \times 10^{-50} \mathrm{sec^{-2}}$. For baryons, assuming local stellar density in our galaxy~\cite{Xue}, we get $M=1 M_{\odot}$, $\Sigma=310~pc^{-2}$ , and $N=1.2 \times 10^{11}$, from which  $\langle{\Delta}^2\rangle \sim 10^{-41}$ and  $\langle\dot{\Delta}^2\rangle \sim 2.24 \times 10^{-64} \mathrm{sec^{-2}}$. Therefore for PSR B1937+21, we obtain $|P - P_0| \sim 4.1 \times 10^{-22}$ sec and $|\dot{P} - \dot{P_0}| \sim 3.6 \times 10^{-28}$. As we can see, these are too small to be observed with current precision. 

Therefore, we conclude that it is not possible to detect any observational signatures of time-dependent modulation of galactic Shapiro delay on the observed period of the 
pulsar with the current technology. However, optimistically  once SKA comes online and the timing residuals are reduced to nanoseconds, one would need  to monitor this pulsar over a baseline of 20-25 years to see at least one dark matter halo transiting near the line of sight between the Earth and the pulsar. Moreover to detect such signatures, one would need to use non-standard 
data analysis methods, where the time-dependent component of galactic Shapiro delay is kept as a free parameter~\cite{Fry}.
The cumulative effects from dark and baryonic matter on the variance of the pulsar period and its derivative are much smaller 
and cannot be detected. Alternately, another strategy to look  for time-dependent Shapiro delay  would be to look at the residual power spectrum over a long period of time  to distinguish it from other sources of noise and signals from a gravitational wave background~\cite{Baghram}, or to look at the auto-correlation function of the  timing residuals~\cite{Kopeikin05}.

\section{Conclusions}
\label{sec:conclusions}

In this paper, we have stressed that the total time required for any astrophysical messenger to reach
the Earth from a given source, includes the Shapiro delay contribution from the total gravitational potential
of our Milky Way galaxy. We provide some examples of current and future astrophysical measurements, where  one-way Shapiro delay plays 
a pivotal role.  We then calculate the static one-way Shapiro delay for PSR B1937+21 (which is the first ever discovered millisecond pulsar), by summing up the contributions from both the dark matter as well as baryonic matter. For the dark matter mass distribution, we have used the  NFW profile parameters from ~\cite{Klypin} and obtain a delay of about 4.74 days. For the baryonic mass, we have used the bulge and disk mass from~\cite{Xue} and get a value of approximately 0.22 days.  Therefore, the total additional delay experienced by the radio signal is about five days. We then estimated the velocity dependent  modulation of  Shapiro delay on the pulsar period by one dark matter halo with mass of 100 $M_{\oplus}$ orbiting near the line of sight between the Earth and the pulsar, to be about five nanoseconds over a five year time-scale. However, realistically one expects 
only one  such transit  in 20 years from all the matter distribution along the line of sight.  We also calculate the variance of the change in pulsar's period and its derivative, due to the cumulative effects of all the matter distribution over shorter time scales, and find that the absolute
value of the difference in the pulsar period is $\mathcal{O}(10^{-22})$ sec. Therefore, the time-dependent modulation to the static one-way Shapiro delay for this pulsar from the matter distribution along the line of sight is currently unobservable. 
In order to detect observational signatures, one would need to reduce the timing residuals to $\mathcal{O}(ns)$, monitor this
pulsar for over a 25 year time-scale, and use sophisticated signal processing techniques with non-standard assumptions. In principle however,
once the uncertainty in distance and dark matter mass distribution is reduced, the static Shapiro delay calculation can be incorporated in {\tt TEMPO2} or other pulsar data analysis packages.

Therefore, the absolute value of the one-way Shapiro delay can never be measured and its time-variation for this millisecond pulsar  is too small to be detected within the next decade. However,  one should not forget that one-way Shapiro delay is ubiquitous and  experienced by all photons, neutrinos, gravitational waves,  and cosmic rays from the gravitational potential of our Galaxy along the line of sight, 
 as they reach the Earth from  distant sources in the Universe. It  will  play a pivotal role   in  multi-messenger astronomy.

\begin{acknowledgements}
We thank S. Kopeikin, M. Longo, and R. Woodard for many  discussions on Shapiro delay over the years. EOK   acknowledges support from Tubitak Grant Number: 112T817.
\end{acknowledgements}
\bibliography{galacticshapiro2}
\end{document}